\documentclass[twocolumn,showpacs,preprintnumbers,amsmath,amssymb]{revtex4}

\usepackage{graphicx}
\usepackage{dcolumn}
\usepackage{bm}

\begin{document}

\preprint{}
\input{epsf.tex}

\title{Quantum Phases of Ultracold Bosonic Atoms in two Bands of an Optical-Lattice coupled by a Cavity Field}

\author{Hashem Zoubi, and Helmut Ritsch}

\affiliation{Institut f\"{u}r Theoretische Physik, Universit\"{a}t Innsbruck, Technikerstrasse 25, A-6020 Innsbruck, Austria}

\date{16 February, 2009}

\begin{abstract}
We study the quantum phase transitions between superfluid and Mott insulator states for ultracold bosons occupying two bands of an optical lattice. The two atomic states are resonantly coupled by a single cavity mode which mediates transitions between the two bosonic particle modes via absorption or emission of a cavity photon. This coupling between the bands shifts the appearance of the Mott insulator phase towards deeper optical lattice potentials and stronger on-site interaction strength, as atomic coherence can build up via photon assisted tunneling in both bands. Varying the intra and interband on-site interactions leads to several different atomic phase configurations. There are even parameter regions where a mean field approach predicts concurrence of a Mott insulator state in one band, while atoms in the second band stay superfluid.
\end{abstract}


\maketitle

\section{Introduction}

Cavity Quantum Electrodynamics (CQED), i.e. the dynamics of an quantized electromagnetic field in a resonator resonantly coupled to a system with quantized energy levels, is a central theoretical thought model to exhibit the physics of coupled quantum dynamics and measurements \cite{Haroche}. Experimental realizations of CQED were e.g. implemented for single Rydberg atoms in a superconducting microwave resonator or for ground state alkali atoms in high-Q optical resonators. In both cases the strong coupling limit could be reached, where light matter coupling dominates the environmental decoherence. Later other implementations using solid state materials or ions followed \cite{Kavokin}. Recently, a Bose-Einstein Condensate (BEC) of ultracold atoms could be trapped within an optical high-Q cavity in the strong resonant coupling regime \cite{Esslinger,Colombe,Slama}. As an important example, the role of cavity induced light forces on the atoms for the case of off-resonant coupling of a BEC in an optical cavity was also widely studied, e.g. in \cite{Horak,Maschler}. It seems just a matter of time that ultracold atoms in an optical lattice within a cavity will be achieved experimentally entering the full quantum many body domain of CQED \cite{Lewenstein}.

In the present work we broaden our recent investigations on excitons and polaritons in ultracold atom optical lattices, where we assumed a frozen Mott insulator state of the atoms in one or both bands \cite{ZoubiA,ZoubiB}. Hence we study the motional quantum dynamics of a two-mode Bose gas in a prescribed optical lattice with resonant coupling of atomic excitations to cavity photons, to see under which conditions such a common Mott state will exist and be stable. In contrary to some recent studies on cavity meditated dipole forces \cite{Maschler}, we assume resonance between the cavity mode and the atomic transition, so that cavity induced light forces are very small compared to the prescribed lattice potentials \cite{Vukics} and will thus be ignored in the following. In such cases collective electronic excitations (excitons) and cavity polaritons \cite{ZoubiA,ZoubiB,HashemA,ZoubiD} depend on the atomic position distribution, which we have demonstrated studying defects in optical lattices \cite{ZoubiC}. Here we will study the stability conditions of such a lattice itself, which was the basis of our previous calculations. While the optical lattice is formed by external classical fields far off resonance to any atomic transitions, the cavity field is represented by quantized single mode close to resonance to an atomic transition. In practise analogous coupling can be achieved by a resonant Raman transition between the two states, where one arm is connected to the cavity field \cite{Salzburger}.

In the following we concentrate on the modified phase diagram for such a system with particular emphasis on the influence of the cavity coupling on the transition from a superfluid to a Mott insulator phase. As we consider two boson modes for the ground and excited state atoms, quantum phase transitions can occur in each band separately or in both bands together. Besides site to site hopping in each band an atom can also jump by changing bands hopping to another site and changing back to the original band. Hence even if the conditions for a Mott insulator are fulfilled in one band, the coupling can induce long range coherence in this band via indirect tunneling. Similar strong on-site repulsion between atoms in different bands could lead to an alternating state of atoms in being at different bands in neighboring sites. While for any concrete setup the hopping amplitudes and on-site interaction strengths can in principle be explicitly calculated, that can be changed by modifying lattice depths and scattering lengths of the particles. Hence here we will keep them as free parameters to explore a wide parameter space of such systems. While as first guess one might think of an optical cavity transition frequency for such a system, also implementation with microwave resonators come to mind, where decay and decoherence are much smaller in general \cite{Jose}.

The paper is organized as follows. In section 2 we review the quantum phase transition for the single-component and two-components Bose-Hubbard model with interaction. In section 3 we add the cavity interaction between the two-components Bose-Hubbard model. As above the corresponding phase-diagram is calculated in the mean field theory in section 4. A summary of the results appears in section 5. The appendix includes the derivation of the second order perturbation theory.

\section{Ultracold atoms in an optical lattice}

\subsection{Single-Component Bose-Hubbard Model}

Bose-condensation of atomic vapors and successive loading of the particles into an optical trap were ground breaking recent experimental achievements, demonstrating unprecedented control of internal and external quantum state of particles. {\it Cold atoms} have been trapped in laser fields forming a standing wave which creates a periodic lattice of microtraps \cite{Jaksch,Greiner}. When loaded into such an {\it optical lattice} cold atoms still hop from site to site via tunneling. In addition, atoms occupying the same lattice site will repel each other due to collisional interactions. The stronger the laser field, the deeper the lattice and the slower the hopping rate of the atoms. At the same time the interaction energy between the atoms becomes stronger, as the atoms on one lattice site get more compressed. Thus laser light allows one to control the kinetic versus the repulsion energy of the quantum gas. In this way one can switch, e.g., a Bose gas from a weakly interacting gas, where atoms form a {\it superfluid}, to a strongly correlated regime, where atoms avoid each other and localize into individual sites forming a so called {\it Mott insulator} phase. This cross over between two quantum phases at $T=0$ as a function of an external parameter is called a {\it quantum phase transition}, a phenomenon which in the meantime has been observed in many laboratories.

Naturally this quantum phase transition from the {\it superfluid} to the {\it Mott insulator} phase was widely studied both theoretically \cite{Jaksch,Zoller,Dalibard} and experimentally \cite{Greiner,Dalibard,Spielman} and it was found that the {\it Bose-Hubbard model} \cite{Fisher} predicts most properties of this phase transition very well. Trapping potentials can be experimentally controlled and using different initial BEC-densities the optical lattice can be filled with {\it one}, {\it two} or even more atoms per site on average by changing the relevant parameters.

Mathematically, the {\it single component} Bose-Hubbard model is represented by
\begin{equation}
H=-J\sum_{\langle i,j\rangle}b_i^{\dagger}b_j+\sum_i(\varepsilon_i-\mu)\ b_i^{\dagger}b_i+\frac{U}{2}\sum_ib_i^{\dagger}b_i^{\dagger}b_ib_i.
\end{equation}
The $i$ summation is over the lattice sites, where $b_i^{\dagger}$ and $b_i$ are the creation and annihilation operators of an atom at site $i$, and $\varepsilon_i$ is the atoms on-site energy, which includes the external trap potential. $\mu$ is the chemical potential accounting for a possible exchange of atoms with an external atom reservoir. We assume here finite hopping (atom transfer) only between {\it nearest} neighbor sites, as indicated by the bracket, with the hopping parameter $J$. Here $U$ is the effective {\it on-site} atom-atom interaction strength, which is taken to be a repulsive and is characterized by the $s$-wave scattering length. Close to $T=0$ the atoms stay localized in the lowest Bloch band of the optical lattice potential. Experimentally the relative magnitude of parameters $J$ and $U$ can be controlled by changing the laser intensity (potential depth). In the limit of $J\gg U$ the atom hopping among the lattice sites dominates the dynamics, and the atomic states are spread out over the whole lattice, so that we get a superfluid phase. In the opposite limit $J\ll U$, the on-site atom-atom interaction dominates and the atoms are localized on individual sites, so that we have a Mott insulator phase. Between the two limits we get a quantum phase transition, which is nicely predicted by the Bose-Hubbard model presented above \cite{Jaksch}. The Mott insulator phase is characterized by a fixed integer number of atoms per site.

In figure (1) we plot the corresponding phase diagram for the plane $(\mu/zJ)-(U/zJ)$ which is scaled by $zJ$, where $z$ is the number of nearest neighbors. The superfluid (SF) phase appears outside the three Mott-insulator (MI) phase regions, which are for one, two, and three atoms per site, that is $n=1,2,3$. Here we assumed $\varepsilon_i=0$. Such a phase diagram can be conveniently calculated in a mean-field approach \cite{Sheshadri,Oosten}.

\begin{figure}[h!]

\centerline{\epsfxsize=7.0cm \epsfbox{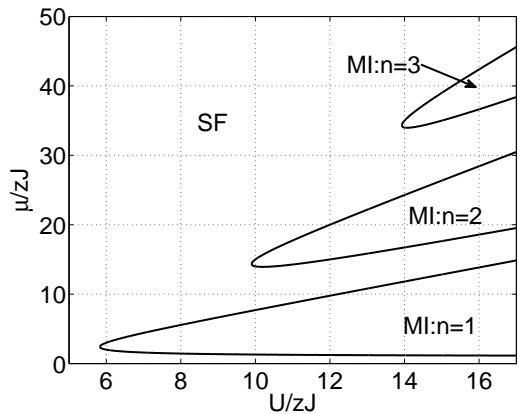}}
\caption{Phase diagram: $(\mu/zJ)$ vs. $(U/zJ)$. The superfluid SF and the three Mott insulator MI regions, for $n=1,2,3$, are shown.}
\end{figure}

\subsection{Two-Components Bose-Hubbard Model}

 Let us extend this model to include two different internal states of the atom, called ground and excited state. Hence we consider a boson gas of two-level atoms described by {\it two kinds} of bosons differing by their internal state. Equivalently also to different bands for atoms in the same internal state could be used. As before, the atoms are cooled and loaded on an optical lattice. In general the light forces of the lattice lasers are acting differently on ground and excited state atoms, and we get different optical lattice potentials for them as shown in figure (2). Quite generally we can, nevertheless, assume optical lattice potentials with minimums at the {\it same} positions. For certain wavelengths, called magic laser wave lengths \cite{Katori}, even identical potentials can be found, which has particular interest for building lattice atom clocks. For cold enough particles the ground and excited state atoms can be assumed to be localized in their first Bloch band. In this case the system is described by a two-component Bose-Hubbard model \cite{Chen,Lukin}, given by the Hamiltonian
\begin{eqnarray}
H&=&-J_g\sum_{\langle i,j\rangle}b_i^{\dagger}b_j-J_e\sum_{\langle i,j\rangle}c_i^{\dagger}c_j \nonumber \\
&+&\frac{U_g}{2}\sum_ib_i^{\dagger}b_i^{\dagger}b_ib_i+\frac{U_e}{2}\sum_ic_i^{\dagger}c_i^{\dagger}c_ic_i+U_{eg}\sum_ib_i^{\dagger}c_i^{\dagger}b_ic_i \nonumber \\
&+&\sum_i(\varepsilon_i^g-\mu_g)\ b_i^{\dagger}b_i+\sum_i(\varepsilon_i^e-\mu_e)\ c_i^{\dagger}c_i,
\end{eqnarray}
where $b_i^{\dagger}$ and $b_i$ are the creation and annihilation operators of a ground state atom at site $i$, respectively, with the on-site energy $\varepsilon_i^g$, and chemical potential $\mu_g$; $c_i^{\dagger}$ and $c_i$ are the creation and annihilation operators of an excited state atom at site $i$, respectively, with on-site energy $\varepsilon_i^e$, and chemical potential $\mu_e$; where $\varepsilon_i^e=\varepsilon_i^g+\hbar\omega_a$, with $\omega_a$ being the effective atomic transition frequency, which can include AC Stark-shifts relative to the free atomic transition frequency. Here $J_g$ and $J_e$ are the hopping parameters of ground and excited state atoms, respectively. $U_g$ and $U_e$ are the on-site ground and excited state atom-atom interactions, respectively, and $U_{eg}$ is the on-site ground-excited atom interaction related to scattering between ground and excited state atoms. This extended model leads to a much richer phase diagram including also boundaries for the superfluid to the Mott insulator quantum phase transition. The Mott insulator phase with a fixed atom number per site can be of different form from above, i.e. we have one atom per site but it does not matter, if it is ground or excited state atom. Also it is possible to get phases with two atoms per site, with a number of options: two ground state atoms, two excited state atoms, or one ground and the other excited state atoms per site, and so on to higher atom number per site.

\begin{figure}[h!]
\centerline{\epsfxsize=6.0cm \epsfbox{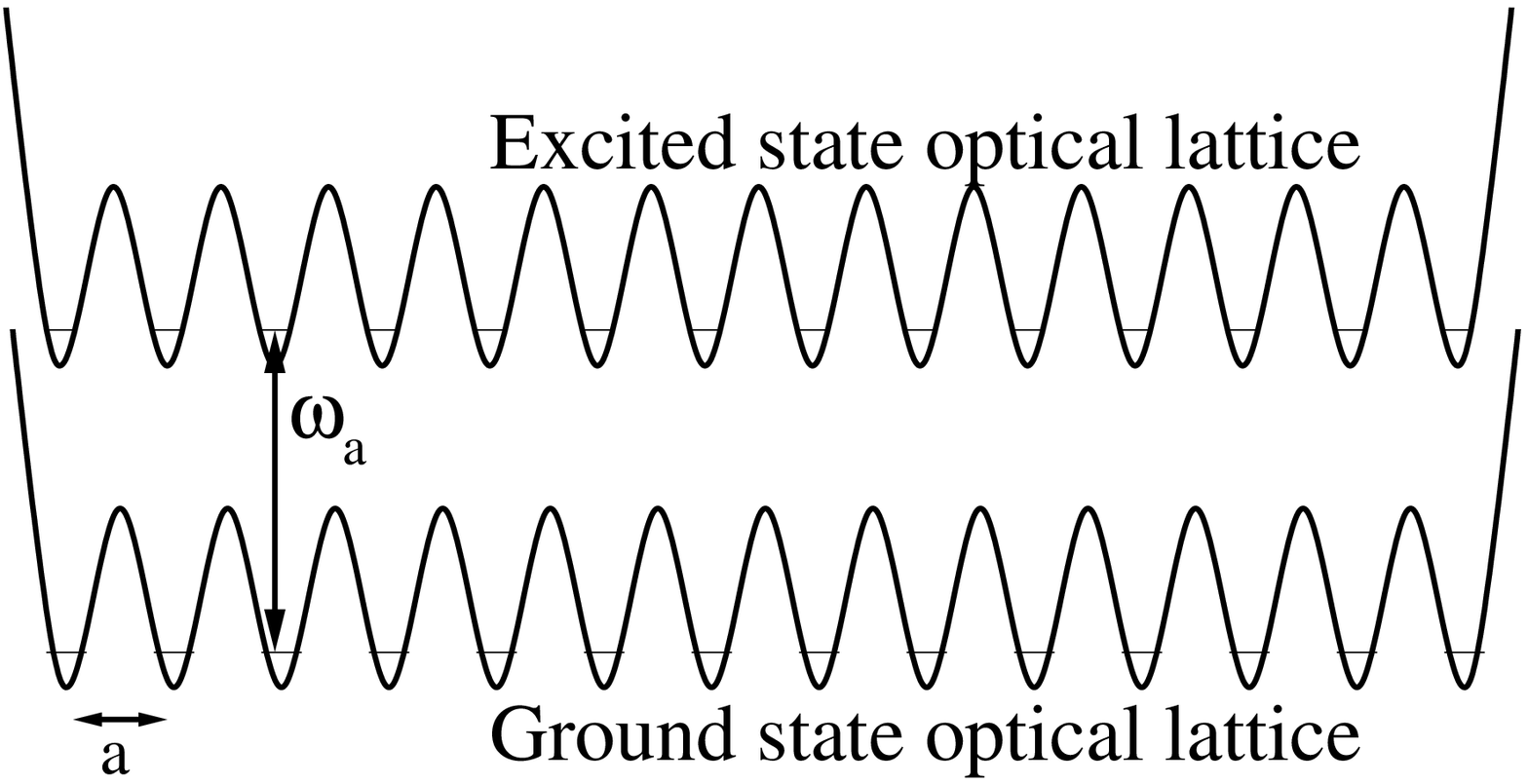}}
\caption{Schematic plot of the ground and excited state optical lattice potentials.}
\end{figure}

Let us show some typical examples now. In figures (3-4) we plot the phase diagram for the plane $(\mu_g/zJ_g)$ vs. $(U_g/zJ_g)$, and $(\mu_e/zJ_e)$ vs. $(U_e/zJ_e)$, which are scaled by $zJ_g$ and $zJ_e$. In figure (3) we used for the scaled atom-atom coupling the number $U_{eg}/zJ_g=U_{eg}/zJ_e=15$. The plot show the superfluid phase (SF) and the Mott insulator phase (MI) for one ground or excited state atom per site, that is $n_g=n_e=1$, where the transition lines for ground and excited atoms are coincide. Figure (4) is with different atom-atom coupling, where $U_{eg}/zJ_g=15$ and $U_{eg}/zJ_e=20$. Here the transition line (Le) is for excited state atoms, and (Lg) for ground state atoms. Beside the superfluid phase (SF), we get three other regions. The Mott insulator (MI) region with one ground state and one excited state atom per site. The (SM) region, where the excited state atoms are in the Mott insulator phase with one atom per site, and the ground state atoms are in the superfluid phase. The (MS) region, where the ground state atoms are in the Mott insulator phase with one atom per site, and the excited state atoms are in the superfluid phase. In the previous two cases of figures (3-4) we assumed $\varepsilon_i^g=\varepsilon_i^e=0$. In figure (5) we used for the scaled atom-atom coupling the number $U_{eg}/zJ_g=U_{eg}/zJ_e=15$, with $\varepsilon_i^g/zJ_g=0$ and $\varepsilon_i^e/zJ_e=100$. Here the excited and ground phase transition lines of figure (3) split. The phase diagram is calculated here by using the mean-field approach \cite{Chen}.

\begin{figure}[h!]
\centerline{\epsfxsize=7.0cm \epsfbox{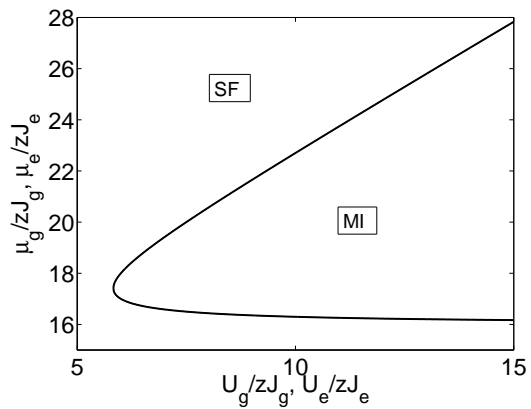}}
\caption{Phase diagram: $(\mu_g/zJ_g)$ vs. $(U_g/zJ_g)$, and $(\mu_e/zJ_e)$ vs. $(U_e/zJ_e)$. The superfluid SF and the Mott insulator MI regions, for $n_g=n_e=1$, are shown. We used $U_{eg}/zJ_g=U_{eg}/zJ_e=15$.}
\end{figure}

\begin{figure}[h!]
\centerline{\epsfxsize=7.0cm \epsfbox{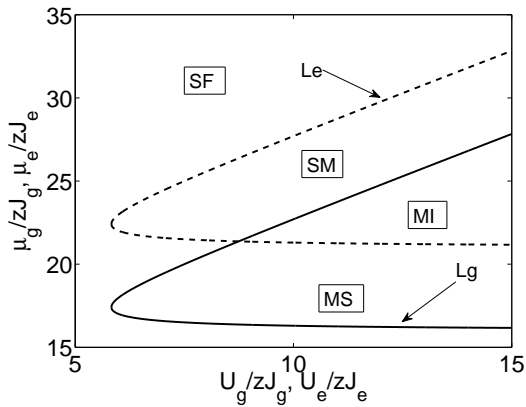}}
\caption{Phase diagram: $(\mu_g/zJ_g)$ vs. $(U_g/zJ_g)$, and $(\mu_e/zJ_e)$ vs. $(U_e/zJ_e)$. The SF and MI regions, for $n_g=n_e=1$, are shown. The SM region is for the Mott insulator of excited atoms and superfluid of ground atoms, and the MS region is for the Mott insulator of ground atoms and superfluid of excited atoms. The full-line Lg is for the transition line of ground atoms, and the dashed-line Le is for the transition line of excited atoms. We used $U_{eg}/zJ_g=15$ and $U_{eg}/zJ_e=20$.}
\end{figure}

\begin{figure}[h!]
\centerline{\epsfxsize=7.0cm \epsfbox{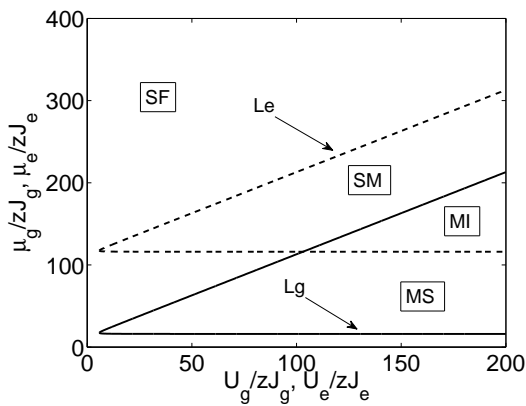}}
\caption{Phase diagram: $(\mu_g/zJ_g)$ vs. $(U_g/zJ_g)$, and $(\mu_e/zJ_e)$ vs. $(U_e/zJ_e)$. The SF, MI, SM, and MS regions, for $n_g=n_e=1$, are shown. We used $U_{eg}/zJ_g=U_{eg}/zJ_e=15$, with $\varepsilon_i^g/zJ_g=0$ and $\varepsilon_i^e/zJ_e=100$.}
\end{figure}

\section{An optical lattice within a cavity}

In a next step we will now add resonant interaction with a cavity mode. Note that trapped ultracold atoms within a cavity were already experimentally studied by a number of experimental groups \cite{Esslinger,Colombe,Slama,Jose}, but as far as we know no optical lattices within a cavity were realized yet. Here we investigate a situation as depicted in figure (6), where we start from the previous system of optical lattice with two state atoms, which we now placed within a cavity \cite{ZoubiA}. We assume that the atomic transition is close to resonance with only a single cavity mode equivalently coupled to all atoms. Hence either mode function is constant over the lattice or we consider very long wavelength transitions (microwaves) \cite{Jose}, where the atoms are within a single wavelength. The mode Hamiltonian is $H_c=\varepsilon_c\ a^{\dagger}a$, where $a^{\dagger}$ and $a$ are the creation and annihilation operators of a cavity mode of energy $\varepsilon_c$, respectively, and where $\varepsilon_c\sim\varepsilon_e-\varepsilon_g$. The coupling between the cavity mode and the atomic transition, in the {\it rotating wave approximation}, is
\begin{equation}
H=\sum_i\left(f_i\ c_i^{\dagger}b_i\ a+f_i^{\ast}\ a^{\dagger}\ b_i^{\dagger}c_i\right),
\end{equation}
where $f_i$ is the coupling parameter, which is taken here to be of the {\it electric dipole interaction}. The first term represents the excitation of an atom from the ground state into the excited state by the absorption of a photon, and the second is for the jump of an excited state atom into the ground state by the emission of a photon.

\begin{figure}[h!]
\centerline{\epsfxsize=6.0cm \epsfbox{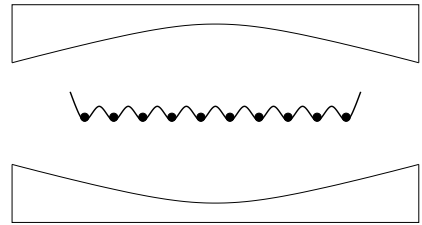}}
\caption{Ultracold atoms in an optical lattice within a cavity.}
\end{figure}

As central goal of the present work we check how the phase diagram is changed as a result of the cavity induced interactions. Here important differences are expected due to the fact that the absorption and emission of cavity photons couple the two kinds of bosons and allow for nonlocal energy transfer between the atoms. A transition from an excited to a lower state atom at one site with photon creation can be compensated by an excitation at any other site by photon absorption. In principle this process can be mediated by a virtual photon, which never actually appears in the mode.
  
In particular we examine, how the superfluid to the Mott insulator phase transition is affected by such coupling to the cavity photons, which should allow to build long range correlations very effectively. At this point our approach here is mainly {\it analytical} based on {\it mean field theory} for atoms and the field. We apply method as it appears in \cite{Sheshadri}, which was successfully used in \cite{Oosten} to one component Bose-Hubbard model, and in \cite{Chen} for two components case, where the superfluid and the Mott insulator phases predicted to a good approximation, and with their quantum phase transition. We assume long life times for both the atomic excitations and the cavity photons, where their line widths are taken to be smaller than the excitation-photon coupling. Hence in the following calculations we neglect the damping for the excitations and the photons. Furthermore, we consider a cavity with a fixed number of photons, which can be achieved by the combination of an external pump and the photon damping.

A two-components Bose-Hubbard model within a cavity in the resonant regime is represented by the Hamiltonian
\begin{eqnarray}
H&=&-J_g\sum_{\langle i,j\rangle}b_j^{\dagger}b_i-J_e\sum_{\langle i,j\rangle}c_j^{\dagger}c_i \nonumber \\
&-&\mu_g\sum_i n_i^g-\mu_e\sum_in_i^e+\varepsilon_c\ n^c \nonumber \\
&+&\frac{U_g}{2}\sum_in_i^g\left(n_i^g-1\right)+\frac{U_e}{2}\sum_in_i^e\left(n_i^e-1\right) \nonumber \\
&+&U_{eg}\sum_in_i^gn_i^e+\sum_i\left(f\ c_i^{\dagger}b_i\ a+f^{\ast}\ a^{\dagger}\ b_i^{\dagger}c_i\right),
\end{eqnarray}
where here $\mu_g=\bar{\mu}_g-\varepsilon^g$ and $\mu_e=\bar{\mu}_e-\varepsilon^e$, with $\bar{\mu}_g$ and $\bar{\mu}_e$ are the ground and excited state atoms chemical potentials, respectively. $\varepsilon^g_i$ and $\varepsilon^e_i$ are taken to be site independent, hence we dropped their index $i$. At pump-cavity mode resonance the chemical potential for cavity photons is zero, that is $\bar{\mu}_c=0$. Furthermore, the photon-excitation coupling $f$ is assumed to be site independent, which is the case for homogeneous optical lattice parallel to the cavity mirrors. We also defined the mean atoms and photons number operator, by $n_i^g=b_i^{\dagger}b_i$, $n_i^e=c_i^{\dagger}c_i$, and $n^c=a^{\dagger}a$.

\section{Phase diagrams in the mean field theory}

In the mean-field theory presented in \cite{Sheshadri} the atoms are subject to the mean-field of the neighboring sites and of the other kind of bosons. Hence the Hamiltonian can be separated into on-site terms and the rest, which can be written as a sum of a free part and an interacting part. Then the interaction part can be treated in perturbation theory.

In the mean-field theory we have
\begin{equation}
\langle b_i\rangle=\langle b_i^{\dagger}\rangle=\sqrt{n_i^g}=\phi_g\ ,\ \langle c_i\rangle=\langle c_i^{\dagger}\rangle=\sqrt{n_i^e}=\phi_e.
\end{equation}
We use the consistent mean-field to obtain the hopping and the excitation-photon coupling terms, by
\begin{eqnarray}
b_j^{\dagger}b_i&=&\left(b_j^{\dagger}+b_i\right)\ \phi_g-\phi_g^2, \nonumber \\
c_j^{\dagger}c_i&=&\left(c_j^{\dagger}+c_i\right)\ \phi_e-\phi_e^2, \nonumber \\
c_i^{\dagger}b_i&=&b_i\ \phi_e+c_i^{\dagger}\ \phi_g-\phi_g\phi_e.
\end{eqnarray}
The Hamiltonian is now given by $H=H_0+H_I$, where
\begin{eqnarray}
H_0&=&-\mu_g\sum_i n_i^g-\mu_e\sum_in_i^e+\varepsilon_c\ n^c+U_{eg}\sum_in_i^gn_i^e \nonumber \\
&+&\frac{U_g}{2}\sum_in_i^g\left(n_i^g-1\right)+\frac{U_e}{2}\sum_in_i^e\left(n_i^e-1\right),
\end{eqnarray}
and
\begin{eqnarray}
H_I&=&-zJ_g\sum_i\left(b_i^{\dagger}+b_i\right)\ \phi_g+zJ_g\sum_i\phi_g^2 \nonumber \\
&-& zJ_e\sum_i\left(c_i^{\dagger}+c_i\right)\ \phi_e+zJ_e\sum_i\phi_e^2 \nonumber \\
&+&\sum_if\ a\left(b_i\ \phi_e+c_i^{\dagger}\ \phi_g-\phi_g\phi_e\right) \nonumber \\
&+&\sum_if^{\ast}\ a^{\dagger}\left(b_i^{\dagger}\ \phi_e+c_i\ \phi_g-\phi_g\phi_e\right).
\end{eqnarray}

The Hamiltonian is on-site separated, where $H_0=\sum_iH_0^i$, with
\begin{eqnarray}
H_0^i&=&-\mu_g\ n_g-\mu_e\ n_e+\varepsilon_c\ n_c+U_{eg}\ n_gn_e \nonumber \\
&+&\frac{U_g}{2}\ n_g\left(n_g-1\right)+\frac{U_e}{2}\ n_e\left(n_e-1\right).
\end{eqnarray}
We formally defined $\bar{n}^c=n^c/N$ in dividing by the number of lattice sites $N$. Hence we dropped the site index, and to write $n_g=n_i^g$, $n_e=n_i^e$, and $n_c=\bar{n}^c$.

The interacting part is $H_I=\sum_iH_I^i$, with $H_I^i=H_{hop}^i+H_{cav}^i$, where, after dropping the index $i$, we get
\begin{eqnarray}
H^i_{hop}&=&-zJ_g\left(b^{\dagger}+b\right)\ \phi_g+zJ_g\ \phi_g^2 \nonumber \\
&-&zJ_e\left(c^{\dagger}+c\right)\ \phi_e+zJ_e\ \phi_e^2,
\end{eqnarray}
and
\begin{eqnarray}
H^i_{cav}&=&f\ a\left(b\ \phi_e+c^{\dagger}\ \phi_g-\phi_g\phi_e\right) \nonumber \\
&+&f^{\ast}\ a^{\dagger}\left(b^{\dagger}\ \phi_e+c\ \phi_g-\phi_g\phi_e\right).
\end{eqnarray}

Now we use perturbation theory, where in the mean-field theory the hopping and the excitation-photon coupling is a perturbation. The system ground state is represented in occupation number states by $|0\rangle=|n_g,n_e,n_c\rangle$. To the zero order we get
\begin{eqnarray} \label{EOO}
\langle 0|H^i_0|0\rangle&=&E_0^{(0)}=-\mu_g\ n_g-\mu_e\ n_e+\varepsilon_c\ n_c+U_{eg}\ n_gn_e \nonumber \\
&+&\frac{U_g}{2}\ n_g\left(n_g-1\right)+\frac{U_e}{2}\ n_e\left(n_e-1\right).
\end{eqnarray}
At ground state we have $\partial E_0^{(0)}/\partial n_e=\partial E_0^{(0)}/\partial n_g=0$, which gives
\begin{eqnarray}
\mu_e&=&U_e\ \left(n_e-\frac{1}{2}\right)+U_{eg}\ n_g, \nonumber \\
\mu_g&=&U_g\ \left(n_g-\frac{1}{2}\right)+U_{eg}\ n_e,
\end{eqnarray}
which are solved to get
\begin{eqnarray}
n_e&=&\frac{U_g\left(2\mu_e+U_e\right)-U_{eg}\left(2\mu_g+U_g\right)}{2\left(U_eU_g-U_{eg}^2\right)}, \nonumber \\
n_g&=&\frac{U_e\left(2\mu_g+U_g\right)-U_{eg}\left(2\mu_e+U_e\right)}{2\left(U_eU_g-U_{eg}^2\right)}.
\end{eqnarray}
Here $n_g$ and $n_e$ need to be integers, hence we define $n_g=n_g^0+\alpha$ and $n_e=n_e^0+\beta$, where $-1/2<(\alpha,\beta)<1/2$, or $(n_g^0-1)<(n_g-1/2)<n_g^0$ and $(n_e^0-1)<(n_e-1/2)<n_e^0$, which is solved for $\mu_g$ and $\mu_e$, and which insure $n_g$ and $n_e$ to be integers. If $n_g$ or $n_e$ is zero, then the ground state is unstable due to the interaction between different atom kinds, $U_{eg}$. At $n_g^0=0$ or $n_e^0=0$, for the inequality to hold we need the conditions $U_g>0$, $U_e>0$, and $U_eU_g>U_{eg}^2$ \cite{Pethick}. Therefore, to avoid instability we assume ground state with non-zero and positive atom number, that is $(n_g,n_g)>0$, or to limit the discussion to the previous stability conditions.

In a second step we are now including higher order corrections to the ground state energy induced by the hopping among nearest neighbor sites and by the excitation coupling to the cavity photons, in calculating higher order terms of the perturbation series. As the first order term is zero $E_0^{(1)}=0$, we thus have to calculate the second order term
\begin{equation} \label{SecondOrder}
E_0^{(2)}=\sum_{n\neq 0}\frac{|\langle 0|H_I|n\rangle|^2}{E_0^{(0)}-E_n^{(0)}}.
\end{equation}
The details of the calculation appear in Appendix A, which yield
\begin{eqnarray}
E^{(2)}&=&\left\{zJ_g+z^2J_g^2\left[\left(\frac{n_g+1}{\mu_g-U_g\ n_g-U_{eg}\ n_e}\right)\right.\right. \nonumber \\
&+&\left.\left.\left(\frac{n_g}{-\mu_g+U_g\ \left(n_g-1\right)+U_{eg}\ n_e}\right)\right]\right. \nonumber \\
&+&\left.|f|^2\left[\left(\frac{\left(n_e+1\right)n_c}{\mu_e+\varepsilon_c-U_e\ n_e-U_{eg}\ n_g}\right)\right.\right. \nonumber \\
&+&\left.\left.\left(\frac{n_e\left(n_c+1\right)}{-\mu_e-\varepsilon_c+U_e\ \left(n_e-1\right)+U_{eg}\ n_g}\right)\right]\right\}\phi_g^2 \nonumber \\
&+&\left\{zJ_e+z^2J_e^2\left[\left(\frac{n_e+1}{\mu_e-U_e\ n_e-U_{eg}\ n_g}\right)\right.\right. \nonumber \\
&+&\left.\left.\left(\frac{n_e}{-\mu_e+U_e\ \left(n_e-1\right)+U_{eg}\ n_g}\right)\right]\right. \nonumber \\
&+&\left.|f|^2\left[\left(\frac{n_gn_c}{-\mu_g+\varepsilon_c+U_g\ \left(n_g-1\right)+U_{eg}\ n_e}\right)\right.\right. \nonumber \\
&+&\left.\left.\left(\frac{\left(n_g+1\right)\left(n_c+1\right)}{\mu_g-\varepsilon_c-U_g\ n_g-U_{eg}\ n_e}\right)\right]\right\}\phi_e^2.
\end{eqnarray}
According to Landau theory, at phase transition the factors of $\phi_g^2$ and $\phi_e^2$ equal to zero, that is
\begin{eqnarray} \label{MixEqs}
&&1+zJ_g\left[\left(\frac{n_g+1}{\mu_g-U_g\ n_g-U_{eg}\ n_e}\right)\right. \nonumber \\
&+&\left.\left(\frac{n_g}{-\mu_g+U_g\ \left(n_g-1\right)+U_{eg}\ n_e}\right)\right] \nonumber \\
&&+\frac{|f|^2}{zJ_g}\left[\left(\frac{\left(n_e+1\right)n_c}{\mu_e+\varepsilon_c-U_e\ n_e-U_{eg}\ n_g}\right)\right. \nonumber \\
&+&\left.\left(\frac{n_e\left(n_c+1\right)}{-\mu_e-\varepsilon_c+U_e\ \left(n_e-1\right)+U_{eg}\ n_g}\right)\right]=0, \nonumber \\
&&1+zJ_e\left[\left(\frac{n_e+1}{\mu_e-U_e\ n_e-U_{eg}\ n_g}\right)\right. \nonumber \\
&+&\left.\left(\frac{n_e}{-\mu_e+U_e\ \left(n_e-1\right)+U_{eg}\ n_g}\right)\right] \nonumber \\
&&+\frac{|f|^2}{zJ_e}\left[\left(\frac{n_gn_c}{-\mu_g+\varepsilon_c+U_g\ \left(n_g-1\right)+U_{eg}\ n_e}\right)\right. \nonumber \\
&+&\left.\left(\frac{\left(n_g+1\right)\left(n_c+1\right)}{\mu_g-\varepsilon_c-U_g\ n_g-U_{eg}\ n_e}\right)\right]=0.
\end{eqnarray}
In the limit $zJ_g,zJ_e\gg|f|^2$ we get
\begin{eqnarray}
&&1+zJ_g\left[\left(\frac{n_g+1}{\mu_g-U_g\ n_g-U_{eg}\ n_e}\right)\right. \nonumber \\
&+&\left.\left(\frac{n_g}{-\mu_g+U_g\ \left(n_g-1\right)+U_{eg}\ n_e}\right)\right], \nonumber \\
&&1+zJ_e\left[\left(\frac{n_e+1}{\mu_e-U_e\ n_e-U_{eg}\ n_g}\right)\right. \nonumber \\
&+&\left.\left(\frac{n_e}{-\mu_e+U_e\ \left(n_e-1\right)+U_{eg}\ n_g}\right)\right],
\end{eqnarray}
which is the result for two-component Bose-Hubbard model in neglected the cavity effect, and which leads to the results in figures (3-5).

In the limit where the excitation-photon coupling is much larger than the hopping, that is $zJ_g,zJ_e\ll |f|^2$, we get
\begin{eqnarray}
&&1+\frac{|f|^2}{zJ_g}\left[\left(\frac{\left(n_e+1\right)n_c}{\mu_e+\varepsilon_c-U_e\ n_e-U_{eg}\ n_g}\right)\right. \nonumber \\
&+&\left.\left(\frac{n_e\left(n_c+1\right)}{-\mu_e-\varepsilon_c+U_e\ \left(n_e-1\right)+U_{eg}\ n_g}\right)\right]=0, \nonumber \\
&&1+\frac{|f|^2}{zJ_e}\left[\left(\frac{n_gn_c}{-\mu_g+\varepsilon_c+U_g\ \left(n_g-1\right)+U_{eg}\ n_e}\right)\right. \nonumber \\
&+&\left.\left(\frac{\left(n_g+1\right)\left(n_c+1\right)}{\mu_g-\varepsilon_c-U_g\ n_g-U_{eg}\ n_e}\right)\right]=0,
\end{eqnarray}
which we are going to examine rigorously. We solve for $\bar{\mu}_g=\mu_g+\varepsilon^g$ and $\bar{\mu}_e=\mu_e+\varepsilon^e$, to get the rescaled results, which are rescaled in dividing by $zJ_g$ and $zJ_e$, and after dropping the bars, we get
\begin{eqnarray} \label{Meg}
\tilde{\mu}_g^{\pm}&=&\frac{\mu_g^{\pm}}{zJ_g}=\left(\tilde{\varepsilon}_g+\tilde{\varepsilon}_c^g\right)+\tilde{L}_g\pm\frac{1}{2}\sqrt{\left(\tilde{G}_g\right)^2-4\tilde{K}_g}, \nonumber \\
\tilde{\mu}_e^{\pm}&=&\frac{\mu_e^{\pm}}{zJ_e}=\left(\tilde{\varepsilon}_e-\tilde{\varepsilon}_c^e\right)+\tilde{L}_e\pm\frac{1}{2}\sqrt{\left(\tilde{G}_e\right)^2-4\tilde{K}_e},
\end{eqnarray}
where
\begin{eqnarray} \label{Eeg}
\tilde{L}_g&=&\frac{L_g}{zJ_g}=\tilde{U}_g\left(n_g-\frac{1}{2}\right)+\tilde{U}_{eg}^g\ n_e-\frac{\tilde{F}}{2}\left(n_g+n_c+1\right), \nonumber \\
\tilde{L}_e&=&\frac{L_e}{zJ_e}=\tilde{U}_e\left(n_e-\frac{1}{2}\right)+\tilde{U}_{eg}^e\ n_g-\frac{\tilde{F}}{2}\left(n_c-n_e\right),
\end{eqnarray}
with
\begin{eqnarray} \label{Geg}
\tilde{G}_g&=&\frac{G_g}{zJ_g}=\tilde{U}_g+\tilde{F}\left(n_g+n_c+1\right), \nonumber \\
\tilde{G}_e&=&\frac{G_e}{zJ_e}=\tilde{U}_e+\tilde{F}\left(n_c-n_e\right),
\end{eqnarray}
and
\begin{eqnarray} \label{Keg}
\tilde{K}_g&=&\frac{K_g}{(zJ_g)^2}=\tilde{F}\tilde{U}_g\left(n_g+1\right)\left(n_c+1\right), \nonumber \\
\tilde{K}_e&=&\frac{K_e}{(zJ_e)^2}=\tilde{F}\tilde{U}_e\left(n_e+1\right)n_c,
\end{eqnarray}
where
\begin{equation} \label{Feg}
\tilde{F}=\frac{|f|^2}{z^2J_gJ_e},
\end{equation}
with $\tilde{U}_g=\frac{U_g}{zJ_g},\ \tilde{U}_e=\frac{U_e}{zJ_e}$, and $\tilde{U}_{eg}^g=\frac{U_{eg}}{zJ_g},\ \tilde{U}_{eg}^e=\frac{U_{eg}}{zJ_e}$, also $\tilde{\varepsilon}_c^g=\frac{\varepsilon_c}{zJ_g},\ \tilde{\varepsilon}_c^e=\frac{\varepsilon_c}{zJ_e}$, and we have $\tilde{\varepsilon}_g=\frac{\varepsilon_g}{zJ_g},\ \tilde{\varepsilon}_e=\frac{\varepsilon_e}{zJ_e}$. As we concentrate in the following on the influence of the coupling to cavity photons, we assume $J_g=J_e=J$, hence we drop the tildes and all the parameters to be considered as scaled in dividing by $zJ$. Furthermore, we assume zero ground state energy, $\varepsilon_g=0$, and the cavity mode is in resonance to the electronic excitation, that is $\varepsilon_e=\varepsilon_c$.

In figure (7) we plot the phase diagram for the plane $\mu_g$ vs. $U_g$, and the plane $\mu_e$ vs. $U_e$. We used for the scaled atom-atom coupling the number $U_{eg}=15$. The plot shows the superfluid phase (SF) and the Mott insulator phase (MI) for one ground or excited state atom per site, that is $n_g=n_e=1$, and where we used for the cavity photons $n_c=1$. The transition line (Le) is for excited state atoms, and (Lg) for ground state atoms. Beside the superfluid phase (SF) and the Mott insulator (MI) regions, we have the (SM) region, where the excited state atoms are in the Mott insulator phase with one atom per site, and the ground state atoms are in the superfluid phase. The (MS) region, where the ground state atoms are in the Mott insulator phase with one atom per site, and the excited state atoms are in the superfluid phase. In the figure we assumed $\varepsilon_c=100$ and for the excitation-photon scaled coupling we used $F=25$. We compare the present results of figure (7) to that of figure (5) for the phase diagram of two-components Bose-Hubbard model without a coupling to cavity photons. We deduce that the effect of the coupling to cavity photons, with one photon per lattice site $n_c=1$, and for strong excitation-photon coupling of $F=25$, is to shift the Mott insulator phase to large atom-atom interactions, which is here one order of magnitude larger. Also we get that the ground state transition line is shifted relative to the excited state one.

In figures (8-11) we plot the scaled chemical potentials $\mu_g$ and $\mu_e$ as a function of different parameters, for $n_g=n_e=1$. In figure (8) the plot is as a function of $U_{eg}$, in using $\varepsilon_c=100$, $F=25$, $U_g=U_e=250$, and $n_c=1$. In figure (9) the plot is as a function of $\varepsilon_c$, in using $U_{eg}=15$, $U_g=U_e=250$, $F=25$, and $n_c=1$. In figure (10) the plot is as a function of $F$, in using $U_{eg}=15$, $U_g=U_e=250$, $\varepsilon_c=100$, and $n_c=1$. It is clear from figure (10) that the Mott insulator phase appears only for a limited range of coupling parameters. In figure (11) the plot is as a function of $n_c$, in using $U_{eg}=15$, $U_g=U_e=250$, $\varepsilon_c=100$, and $F=25$. Also here the Mott insulator phase is obtained for a limited range of mean cavity photon number per site, which ranges from few cavity photons up to about one photon per lattice site.

\begin{figure}[h!]
\centerline{\epsfxsize=7.0cm \epsfbox{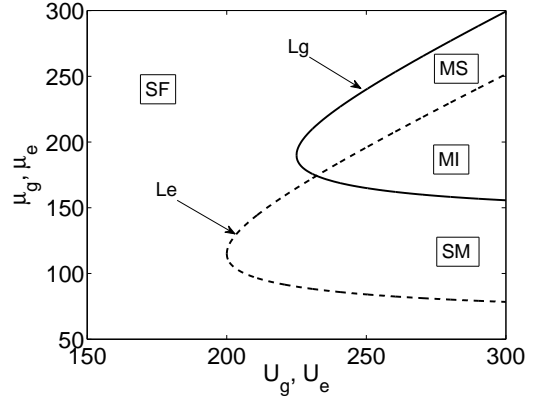}}
\caption{Scaled phase diagram: $\mu_g$ vs. $U_g$, and $\mu_e$ vs. $U_e$, for $n_g=n_e=1$. We have $\varepsilon_c=100$, $F=25$, $U_{eg}=15$, and $n_c=1$. The dashed line is for the excited state atoms, and the full line is for the ground state atoms.}
\end{figure}

\begin{figure}[h!]
\centerline{\epsfxsize=7.0cm \epsfbox{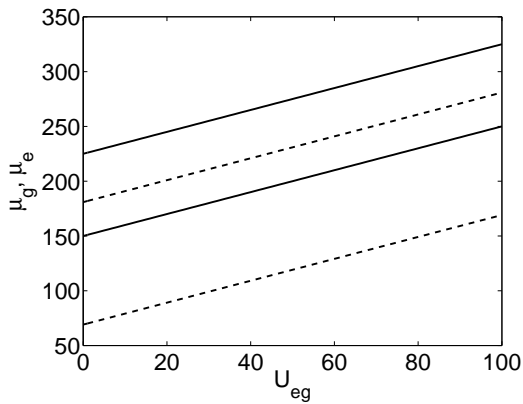}}
\caption{Scaled phase diagram: $\mu_g$ and $\mu_e$ vs. $U_{eg}$, for $n_g=n_e=1$. We have $\varepsilon_c=100$, $F=25$, $U_g=U_e=250$, and $n_c=1$. The dashed lines are for the excited state atoms, and the full lines are for the ground state atoms.}
\end{figure}

\begin{figure}[h!]
\centerline{\epsfxsize=7.0cm \epsfbox{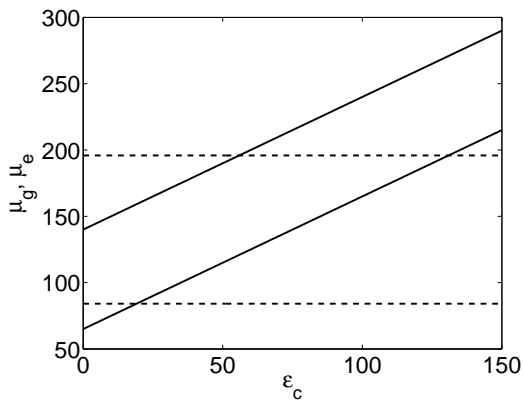}}
\caption{Scaled phase diagram: $\mu_g$ and $\mu_e$ vs. $\varepsilon_c$, for $n_g=n_e=1$. We have $U_{eg}=15$, $U_g=U_e=250$, $F=25$, and $n_c=1$. The dashed lines are for the excited state atoms, and the full lines are for the ground state atoms.}
\end{figure}

\begin{figure}[h!]
\centerline{\epsfxsize=7.0cm \epsfbox{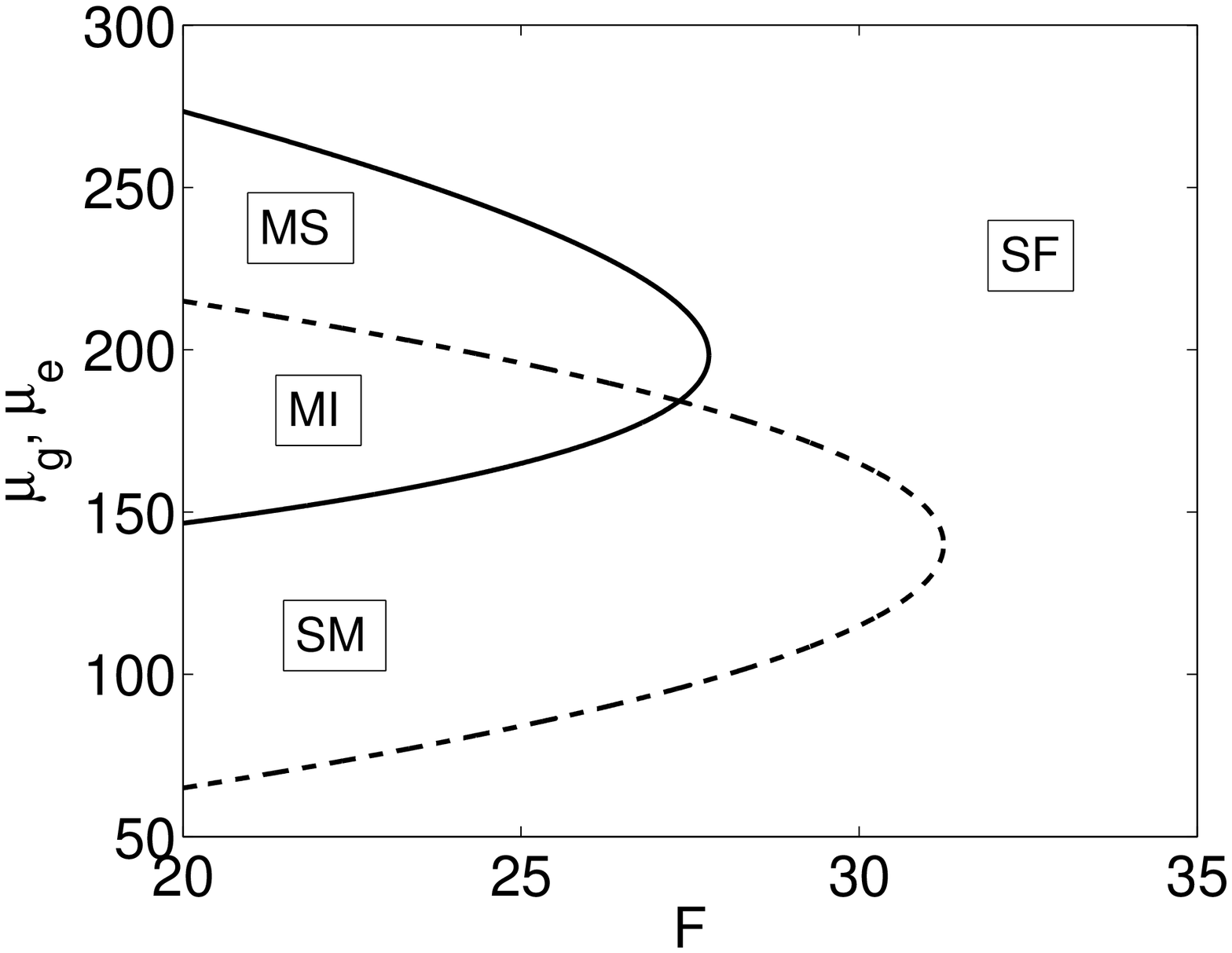}}
\caption{Scaled phase diagram: $(\mu_g,\mu_e)$ vs. $F$, for $n_g=n_e=1$. We have $U_{eg}=15$, $U_g=U_e=250$, $\varepsilon_c=100$, and $n_c=1$. The dashed line is for the excited state atoms, and the full line is for the ground state atoms.}
\end{figure}

\begin{figure}[h!]
\centerline{\epsfxsize=7.0cm \epsfbox{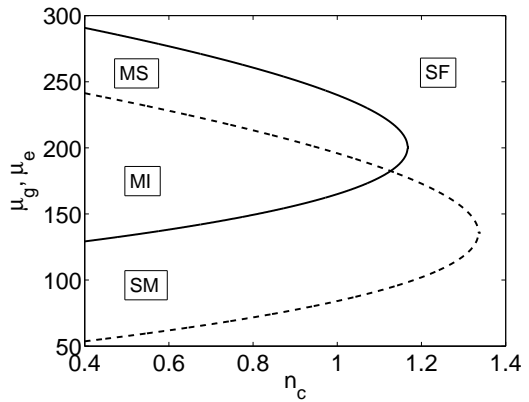}}
\caption{Scaled phase diagram: $(\mu_g,\mu_e)$ vs. $n_c$, for $n_g=n_e=1$. We have $U_{eg}=15$, $U_g=U_e=250$, $\varepsilon_c=100$, and $F=25$. The dashed line is for the excited state atoms, and the full line is for the ground state atoms.}
\end{figure}

In figure (12), beside the lines for the ground and excited state transitions from the superfluid into the Mott insulator with one atom per site (the the upper full line is for the ground state atoms, and the lower full line is for the excited state atoms), we plot the transition line for the case of two ground state atoms per site (the upper dashed line), when the excited state atoms are with zero atoms per site. The plot is for $\mu_g$ vs. $U_g$, and $\mu_e$ vs. $U_e$. We used the parameters: $\varepsilon_c=200$, $F=25$, and $U_{eg}=50$, where the plot is for the two cases of $n_g=n_e=1$ and $n_g=2,\ n_e=0$. Here we assumed $n_c=1$. At these parameters the Mott insulator for $n_g=n_e=1$ is separated from that for $n_g=2,\ n_e=0$, as no crossing appears between the full and the dashed lines. In figure (13) the plot is for the same parameters but now with $U_{eg}=500$. It is clear that for large $U_{eg}$ the transition lines of $n_g=n_e=1$ and $n_g=2,\ n_e=0$ are crossed, and we get correlations between the Mott insulator with one ground and one excited atom per site and that with two ground and zero excited atoms per site. Figures (14-15) are for the transitions with $n_g=n_e=1$ and $n_g=1,\ n_e=0$. In figure (14) we used $U_{eg}=500$, and in figure (15) $U_{eg}=50$, where the other parameters are the same as before. For small $U_{eg}$ we get strong correlations between the Mott insulator phase with $n_g=n_e=1$ and $n_g=1,\ n_e=0$. For larger $U_{eg}$ the correlation appears only for large $U_g$ and $U_e$, but for small ones the transition lines are separated. In figure (16) the plot is with $U_{eg}=250$. Here the two full lines are for the transitions with $n_g=n_e=1$, the lower dashed line is for $n_g=1,\ n_e=0$, and the upper dashed line is for $n_g=2,\ n_e=0$. We conclude that no crossing between the transition lines of $n_g=1,\ n_e=0$ and $n_g=2,\ n_e=0$ is obtained. Furthermore, in figure (17) we plot the phase diagram for the three transitions with $(n_g=n_e=1)$, $(n_g=1,\ n_e=0)$, and $(n_g=0,\ n_e=1)$. As before the two full lines are for the transition with $n_g=n_e=1$, where the upper for the ground state atoms and the lower for the excited state atoms. The upper dashed line for the ground state atoms with $n_g=1,\ n_e=0$, and the lower dashed line for the excited state atoms with $n_g=0,\ n_e=1$. Here we used also $\varepsilon_c=100$, $F=25$, and $U_{eg}=30$. Each zone in the diagram can be explained as before, just to note here the correlations between the different Mott insulator regions. For illustration in figure (18), for the same parameters, we plot the phase diagram for the three transitions with $(n_g=n_e=1)$, $(n_g=2,\ n_e=0)$, and $(n_g=0,\ n_e=2)$. The two full lines are for the transition with $n_g=n_e=1$. The upper dashed line for the ground state atoms with $n_g=2,\ n_e=0$, and the lower dashed line for the excited state atoms with $n_g=0,\ n_e=2$.

\begin{figure}[h!]
\centerline{\epsfxsize=7.0cm \epsfbox{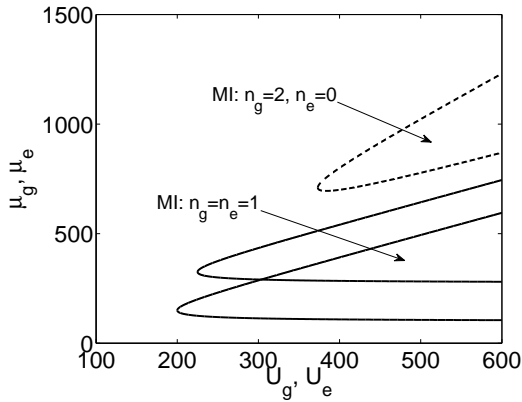}}
\caption{Scaled phase diagram: $\mu_g$ vs. $U_g$, and $\mu_e$ vs. $U_e$, for $n_g=n_e=1$ and $n_g=2,\ n_e=0$. We have $\varepsilon_c=200$, $F=25$, $U_{eg}=50$, and $n_c=1$. The dashed line is for the transition with $n_g=2,\ n_e=0$, and the full lines are for the transition with $n_g=n_e=1$.}
\end{figure}

\begin{figure}[h!]
\centerline{\epsfxsize=7.0cm \epsfbox{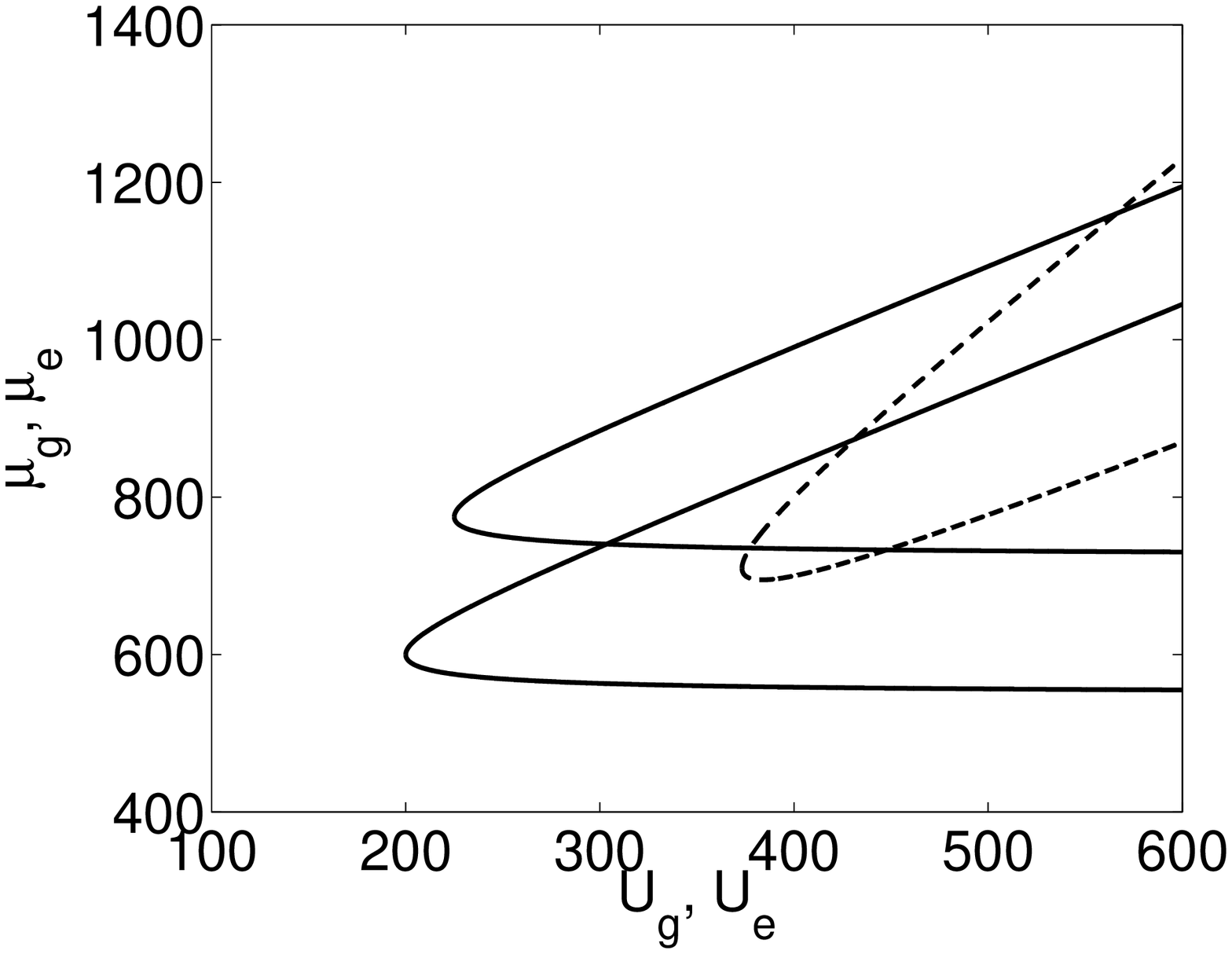}}
\caption{Scaled phase diagram: $\mu_g$ vs. $U_g$, and $\mu_e$ vs. $U_e$, for $n_g=n_e=1$ and $n_g=2,\ n_e=0$. We have $\varepsilon_c=200$, $F=25$, $U_{eg}=500$, and $n_c=1$. The dashed line is for the transition with $n_g=2,\ n_e=0$, and the full lines are for the transition with $n_g=n_e=1$.}
\end{figure}

\begin{figure}[h!]
\centerline{\epsfxsize=7.0cm \epsfbox{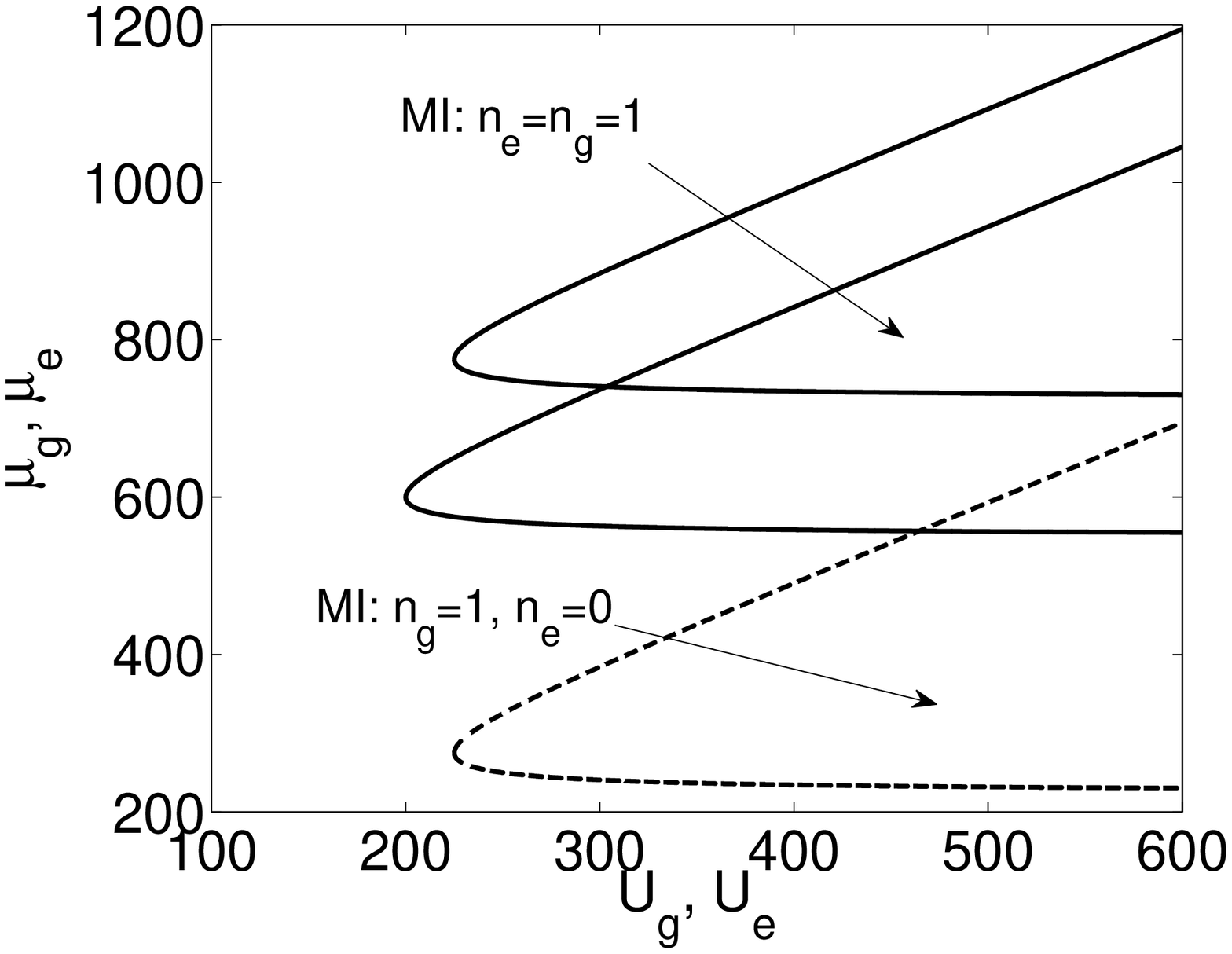}}
\caption{Scaled phase diagram: $\mu_g$ vs. $U_g$, and $\mu_e$ vs. $U_e$, for $n_g=n_e=1$ and $n_g=1,\ n_e=0$. We have $\varepsilon_c=200$, $F=25$, $U_{eg}=500$, and $n_c=1$. The dashed line is for the transition with $n_g=1,\ n_e=0$, and the full lines are for the transition with $n_g=n_e=1$.}
\end{figure}

\begin{figure}[h!]
\centerline{\epsfxsize=7.0cm \epsfbox{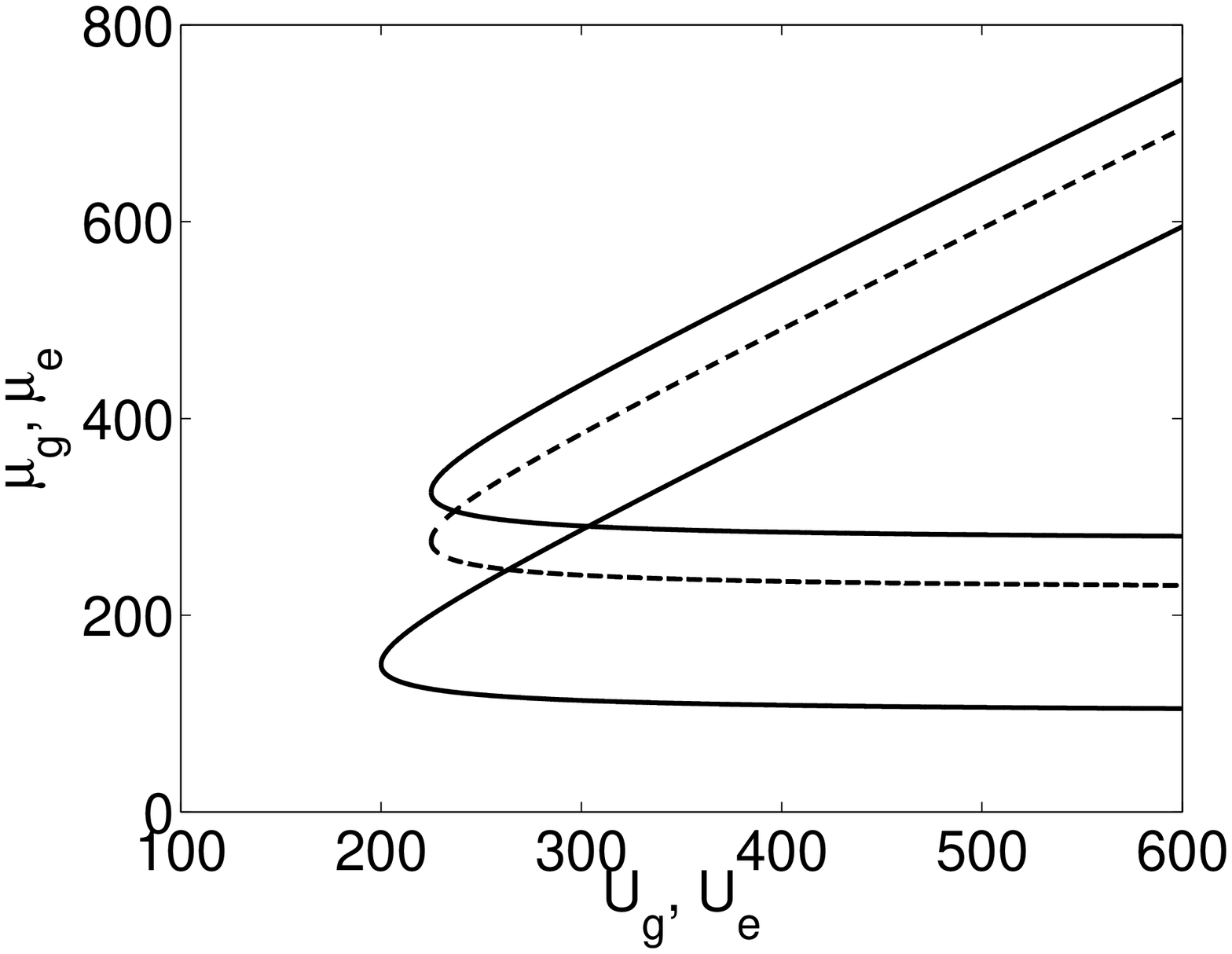}}
\caption{Scaled phase diagram: $\mu_g$ vs. $U_g$, and $\mu_e$ vs. $U_e$, for $n_g=n_e=1$ and $n_g=1,\ n_e=0$. We have $\varepsilon_c=200$, $F=25$, $U_{eg}=50$, and $n_c=1$. The dashed line is for the transition with $n_g=1,\ n_e=0$, and the full lines are for the transition with $n_g=n_e=1$.}
\end{figure}

\begin{figure}[h!]
\centerline{\epsfxsize=7.0cm \epsfbox{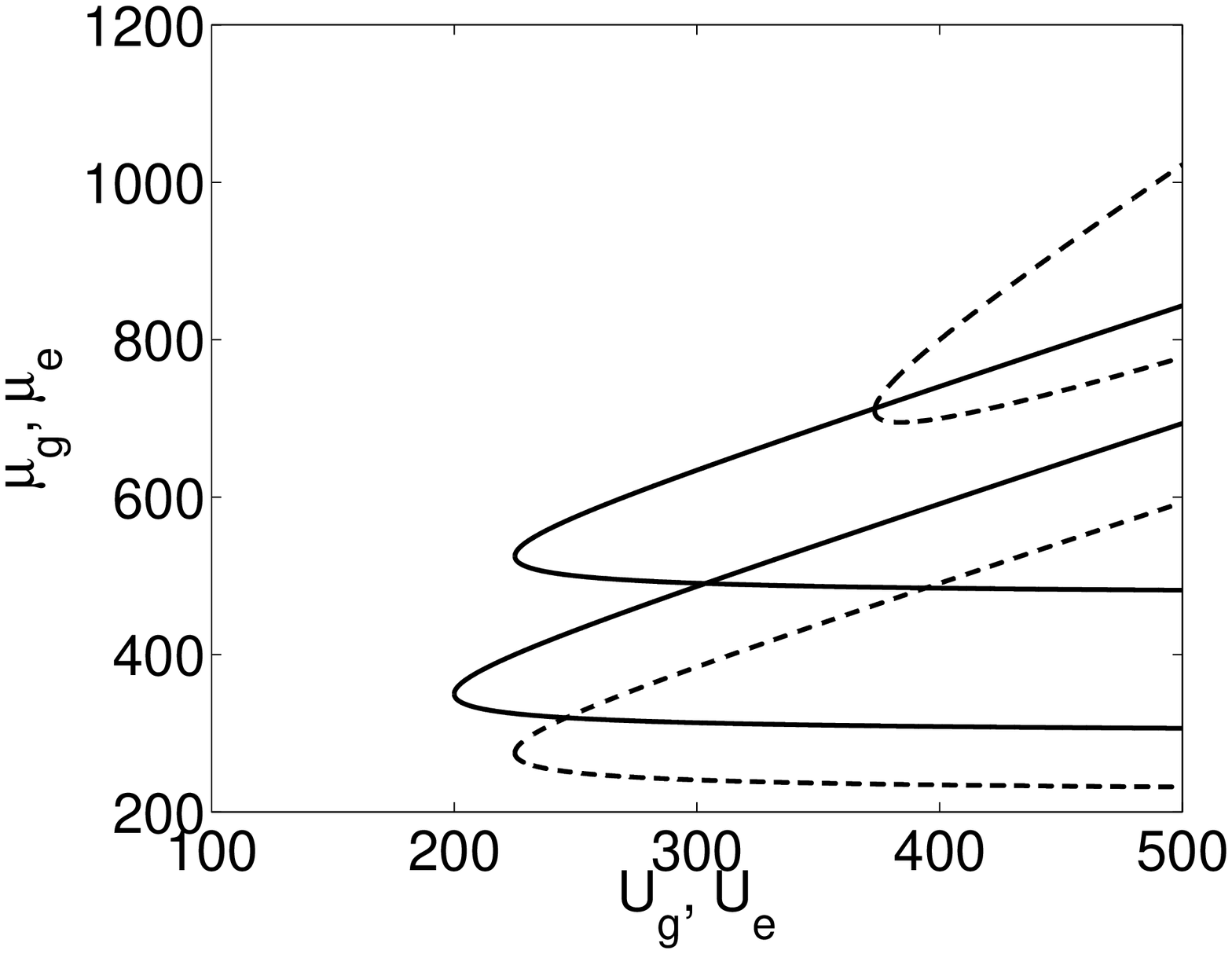}}
\caption{Scaled phase diagram: $\mu_g$ vs. $U_g$, and $\mu_e$ vs. $U_e$, for $(n_g=n_e=1)$, $(n_g=1,\ n_e=0)$, and $(n_g=2,\ n_e=0)$. We have $\varepsilon_c=200$, $F=25$, $U_{eg}=250$, and $n_c=1$. The upper dashed line is for the transition with $n_g=2,\ n_e=0$, the lower dashed line is for the transition with $n_g=1,\ n_e=0$, and the full lines are for the transition with $n_g=n_e=1$.}
\end{figure}

\begin{figure}[h!]
\centerline{\epsfxsize=7.0cm \epsfbox{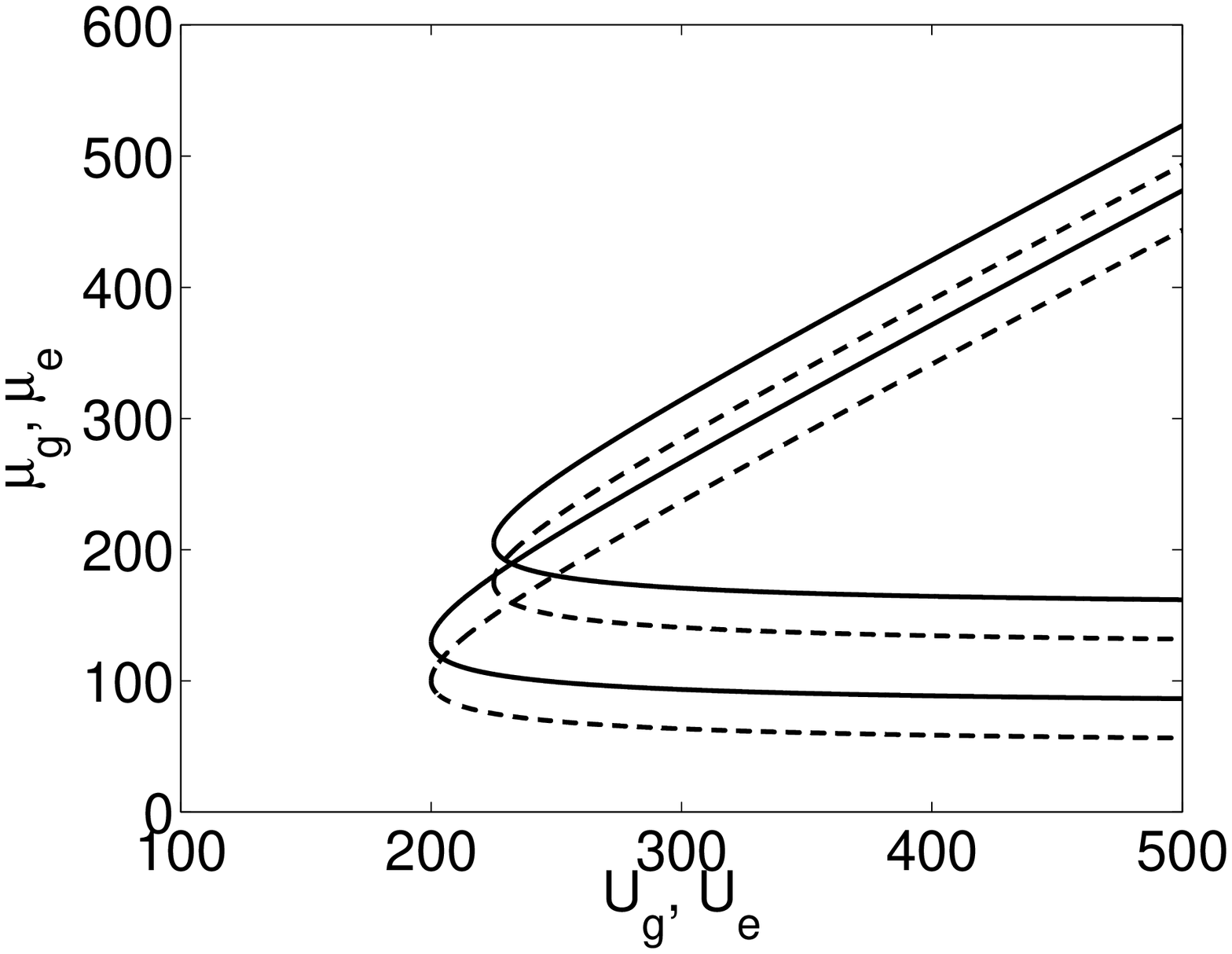}}
\caption{Scaled phase diagram: $\mu_g$ vs. $U_g$, and $\mu_e$ vs. $U_e$, for $(n_g=n_e=1)$, $(n_g=1,\ n_e=0)$, and $(n_g=0,\ n_e=1)$. We have $\varepsilon_c=100$, $F=25$, $U_{eg}=30$, and $n_c=1$. The upper dashed line is for the transition with $n_g=1,\ n_e=0$, the lower dashed line is for the transition with $n_g=0,\ n_e=1$, and the full lines are for the transition with $n_g=n_e=1$.}
\end{figure}

\begin{figure}[h!]
\centerline{\epsfxsize=7.0cm \epsfbox{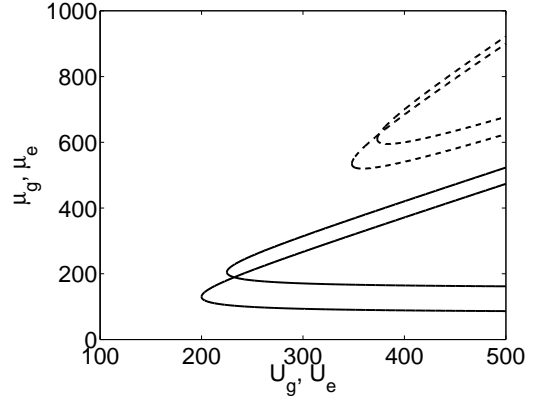}}
\caption{Scaled phase diagram: $\mu_g$ vs. $U_g$, and $\mu_e$ vs. $U_e$, for $(n_g=n_e=1)$, $(n_g=2,\ n_e=0)$, and $(n_g=0,\ n_e=2)$. We have $\varepsilon_c=100$, $F=25$, $U_{eg}=30$, and $n_c=1$. The upper dashed line is for the transition with $n_g=2,\ n_e=0$, the lower dashed line is for the transition with $n_g=0,\ n_e=2$, and the full lines are for the transition with $n_g=n_e=1$.}
\end{figure}

\section{Summary}

We calculated the quantum phase diagram of a dilute gas of ultracold atoms occupying two bands of an optical lattice within a cavity. The atoms in the ground and excited internal states are treated as two kinds of bosons modeled by a two-component Bose-Hubbard Hamiltonian. Using a well established mean field based approach we analytically calculate and plot the parameter conditions for the transition from the superfluid to the Mott insulator phase for a wide range of parameters. When the optical lattice is located within a cavity with a single cavity mode close to resonance with the transition between the two internal electronic states, the possibilities for tunneling of the atoms increase. Hence, recalculating the phase diagram in the strong coupling regime we found that the effect of coupling to cavity photons indeed shifts the Mott insulator region of the phase diagram towards much larger atom-atom interaction parameters. This result is due to the fact that cavity photons tend to delocalize the excited atoms in the optical lattice. Besides the phase transition borderline for one atom per site, a more general approach allows to obtain transition lines for higher numbers of ground and excited state atoms per site showing a rather complex and overlapping general phase diagram.

Nevertheless, for strong enough interactions despite all these couplings, a parameter region of a Mott insulator in both bands, where the atoms are not moving can be identified. Such a state was assumed as basis in our previous calculations on excitons and cavity polaritons in optical lattices. These models now could also be extended to mixed lattice phases.

The results of the present paper should guide future experiments on an optical lattice within a cavity as they are currently set up in a number of laboratories worldwide. We give a qualitative overview and even first quantitative estimates for required system parameters in order to achieve a quantum phase transition towards a perfectly ordered state. Quantitatively our results can be improved in calculating higher order terms of the perturbation theory and in going beyond the mean field theory. Interestingly some preliminary numerical calculations on small system sizes show already completely different possibilities of atomic order like e.g. alternating occupation probabilities between neighboring sites. These phases cannot be captured by our current approach and thus leave much room for future work. Similarly the inclusion of atomic dipole-dipole interactions in form of excitons will also lead to new physics beyond the Bose Hubbard model. 

\begin{acknowledgments}
The work was supported by the Austrian Science Fund (FWF), via the projects P21101 and F4013.
\end{acknowledgments}

\appendix

\section{The second order perturbation theory}

Here we calculate the second order term of the perturbation series Eq.(\ref{SecondOrder}). We start to calculate the matrix elements $\langle 0|H_I|n\rangle$, namely we want to calculate the matrix elements of the operators: $b,\ b^{\dagger},\ c,\ c^{\dagger},\ a,\ a^{\dagger},\ ab,\ a^{\dagger}c,\ ac^{\dagger},\ a^{\dagger}b^{\dagger}$, where $|0\rangle=|n_g,n_e,n_c\rangle$. They have non-vanishing matrix elements between the states
\begin{eqnarray}
|1\rangle=|n_g+1,n_e,n_c\rangle &,& |2\rangle=|n_g-1,n_e,n_c\rangle, \nonumber \\
|3\rangle=|n_g,n_e+1,n_c\rangle &,& |4\rangle=|n_g,n_e-1,n_c\rangle, \nonumber \\
|5\rangle=|n_g,n_e,n_c+1\rangle &,& |6\rangle=|n_g,n_e,n_c-1\rangle, \nonumber \\
|7\rangle=|n_g+1,n_e,n_c+1\rangle &,& |8\rangle=|n_g,n_e+1,n_c-1\rangle, \nonumber \\
|9\rangle=|n_g,n_e-1,n_c+1\rangle &,& |10\rangle=|n_g-1,n_e,n_c-1\rangle, \nonumber \\
\end{eqnarray}
with the energy differences
\begin{eqnarray}
E_0^{(0)}-E_1^{(0)}&=&\mu_g-U_g\ n_g-U_{eg}\ n_e, \nonumber \\
E_0^{(0)}-E_2^{(0)}&=&-\mu_g+U_g\ \left(n_g-1\right)+U_{eg}\ n_e, \nonumber \\
E_0^{(0)}-E_3^{(0)}&=&\mu_e-U_e\ n_e-U_{eg}\ n_g, \nonumber \\
E_0^{(0)}-E_4^{(0)}&=&-\mu_e+U_e\ \left(n_e-1\right)+U_{eg}\ n_g, \nonumber \\
E_0^{(0)}-E_5^{(0)}&=&-\varepsilon_c, \nonumber \\
E_0^{(0)}-E_6^{(0)}&=&\varepsilon_c, \nonumber \\
E_0^{(0)}-E_7^{(0)}&=&\mu_g-\varepsilon_c-U_g\ n_g-U_{eg}\ n_e, \nonumber \\
E_0^{(0)}-E_8^{(0)}&=&\mu_e+\varepsilon_c-U_e\ n_e-U_{eg}\ n_g, \nonumber \\
E_0^{(0)}-E_9^{(0)}&=&-\mu_e-\varepsilon_c+U_e\ \left(n_e-1\right)+U_{eg}\ n_g, \nonumber \\
E_0^{(0)}-E_{10}^{(0)}&=&-\mu_g+\varepsilon_c+U_g\ \left(n_g-1\right)+U_{eg}\ n_e, \nonumber \\
\end{eqnarray}
where $E_0^{(0)}$ is defined in Eq.(\ref{EOO}). The matrix elements are
\begin{eqnarray}
\langle 0|b|1\rangle=\sqrt{n_g+1} &,& \langle 0|b^{\dagger}|2\rangle=\sqrt{n_g}, \nonumber \\
\langle 0|c|3\rangle=\sqrt{n_e+1} &,& \langle 0|c^{\dagger}|4\rangle=\sqrt{n_e}, \nonumber \\
\langle 0|a|5\rangle=\sqrt{n_c+1} &,& \langle 0|a^{\dagger}|6\rangle=\sqrt{n_c}, \nonumber \\
\langle 0|ba|7\rangle=\sqrt{n_g+1}\sqrt{n_c+1} &,& \langle 0|ca^{\dagger}|8\rangle=\sqrt{n_e+1}\sqrt{n_c}, \nonumber \\
\langle 0|ac^{\dagger}|9\rangle=\sqrt{n_e}\sqrt{n_c+1} &,& \langle 0|a^{\dagger}b^{\dagger}|10\rangle=\sqrt{n_g}\sqrt{n_c}. \nonumber \\
\end{eqnarray}
The second order correction for the energy are
\begin{eqnarray}
E^{(2)}_{hop}&=&z^2J_g^2\ \phi_g^2\left\{\left(\frac{n_g+1}{\mu_g-U_g\ n_g-U_{eg}\ n_e}\right)\right. \nonumber \\
&+&\left.\left(\frac{n_g}{-\mu_g+U_g\ \left(n_g-1\right)+U_{eg}\ n_e}\right)\right\} \nonumber \\
&+&z^2J_e^2\ \phi_e^2\left\{\left(\frac{n_e+1}{\mu_e-U_e\ n_e-U_{eg}\ n_g}\right)\right. \nonumber \\
&+&\left.\left(\frac{n_e}{-\mu_e+U_e\ \left(n_e-1\right)+U_{eg}\ n_g}\right)\right\} \nonumber \\
&+&zJ_g\ \phi_g^2+zJ_e\ \phi_e^2,
\end{eqnarray}
and
\begin{eqnarray}
E^{(2)}_{cav}&=&|f|^2\left\{\left(\frac{\left(n_e+1\right)n_c}{\mu_e+\varepsilon_c-U_e\ n_e-U_{eg}\ n_g}\right)\phi_g^2\right. \nonumber \\
&+&\left.\left(\frac{n_gn_c}{-\mu_g+\varepsilon_c+U_g\ \left(n_g-1\right)+U_{eg}\ n_e}\right)\phi_e^2\right. \nonumber \\
&+&\left.\left(\frac{\left(n_g+1\right)\left(n_c+1\right)}{\mu_g-\varepsilon_c-U_g\ n_g-U_{eg}\ n_e}\right)\phi_e^2\right. \nonumber \\
&+&\left.\left(\frac{n_e\left(n_c+1\right)}{-\mu_e-\varepsilon_c+U_e\ \left(n_e-1\right)+U_{eg}\ n_g}\right)\phi_g^2\right. \nonumber \\
&+&\left.\frac{n_c}{\varepsilon_c}\phi_e^2\phi_g^2-\frac{\left(n_c+1\right)}{\varepsilon_c}\phi_e^2\phi_g^2\right\}.
\end{eqnarray}
The terms of the order $\phi_e^2\phi_g^2$ can be neglected at the second order.

\end{document}